# Reduction of Quantum Noise and Increase of Amplification of Gravitational Waves Signals in Michelson Interferometer by the Use of Squeezed States

Y.Ben-Aryeh

Physics Department , Technion-Israel Institute of Technology , Haifa  32000, Israel

e-mail: phr65yb@physics.technion.ac.il

**Abstract**

It is shown in the present Letter that the quantum noise due to high laser intensities in Michelson interferometer for gravitational waves detection can be reduced by sending squeezed vacuum states to the 'dark' port of the interferometer. The experimental details of such physical system have been described in a recent article by Barak and Ben-Aryeh (JOSA-B, **25,** 361(2008)). In another very recent article by Voronov and Weyrauch  (Phys. Rev.A **81**, 053816 (2010)) they have  followed our methods for treating  the same physical system, and have pointed out an error in the sign of  one of our expressions thus claiming for the  elimination of our physical results. I show here a method by which the  expectation value for the photon number operator $\langle \hat{n} \rangle$ is increased and at the same time the standard deviation $(\Delta n)^2$ is reduced. Although due to the mistake in sign in our expression the physical method for obtaining such result is different from our original study [1], the main physical effect can remain .

In the presen Letter I would like to discuss physical consequences following from the analysis made in our our previous study [1] and from that made in [2]. In both references the analysis concentrated about the case in which a squeezed vacuum state is injected into one input port and a very strong coherent state in the other input port of Michelson interferometer. The authors of Ref. 2 have obtained in Eq. (14) of their paper the result for the output of the interferometer under special conditions (defined by $\gamma = (\pi/2) + \delta$, where the parameter $\gamma$ parametrized the splitting ratio to the incoming beam splitter (BS) in port 1 and $\delta$ is an additional very small phase shift):

$$|\psi\rangle_{OUT} = \hat{D}_1\left(-\alpha\delta(1-\cosh(r))\right)\hat{D}_1\left(-\left(\alpha^*\delta\right)\exp(i\theta)\sinh(r)\right)\hat{S}_1(\varsigma)\hat{D}_1(-\alpha\delta)\hat{D}_2(\alpha)|00\rangle \quad (1)$$

Here $\alpha = |\alpha|exp(i\phi)$ is the coherent parameter, $\varsigma = r\exp(i\theta)$ is the squeezing parameter of the input squeezed vacuum state, $\delta$ is a small phase shift introduced by a small external perturbation (e.g. by a gravitational wave), $\hat{S}_1$ is the squeezing operator, $\hat{D}_1$ and $\hat{D}_2$ are coherent displacement operators, $\theta$ and $\phi$ are, respectively, the phases of the coherent and squeezing operator.

Equation (1) is almost equivalent to our previous equation [Eq. (79) in Ref. 1], but the first displacement operator in this equation has a negative sign which is different from a positive sign presented by us. Therefore the physical conclusions presented by them [2] following Eq. (1) are different from us, and they have claimed that the physical effects presented by us are eliminated. Although their correction in the sign in Eq. (1) is right I would like to show here that a basic physical effect is not eliminated.

There is a certain mistake in the derivation of their equations (15-16) [2], since they assume a factor $\Delta(r,\theta,\phi)$ to be real but I find it to be complex. I will use, therefore, in the following analysis the corrected Eq. (1). But before making this analysis let me give a quite simple physical interpretation of Eq. (1). According to the analysis given in Refs. 1 and 2 in the second output of the interferometer we obtain approximately a coherent state given by $\hat{D}_2(\alpha)|0\rangle_2$. Since the input coherent state assumed in the present work is very strong ($|\alpha|$ might be larger than $10^8$ for gravitational waves detection) it exits under the present conditions in output port 2 approximately unchanged by the squeezed vacuum state. In the output 1 we should notice the changes made by the squeezed vacuum state. When a vacuum state enters in the 'dark' input port, a small perturbation represented by a small phase shift $\delta$ will induce a coherent state exiting the 'dark' output port with a coherent parameter absolute value $|\alpha\delta|$. Such relation is a special case of Eq. (1) since for input vacuum state we can assume in Eq. (1): $\cosh(r) = 1$; $\sinh(r) = 0$; $\hat{S}_1(\varsigma) = 1$ and then Eq. (1) is reduced to that of input vacuum state.

Eq. (1) can be rearranged by using well known equations for displacement operators and squeezed states (e.g. Eqs. C1 in [2]). Multiplication of the two first displacement operators of Eq. (1) can be replaced as

$$\hat{D}_1[-\alpha\delta(1-\cosh(r))]\hat{D}_1[-\alpha^*\delta\exp(i\theta)\sinh(r)] = \hat{D}_1\left[\alpha\delta\{\cosh(r)-1-\sinh(r)\exp(i(\theta-2\phi))\}\right] \quad (2)$$

where an additional phase factors which do not affect photon statistics photon numbers are neglected.

The interesting case occurs under the condition $\theta - 2\phi = 0$. Then we get:

$$\hat{D}_1\left[\alpha\delta\{\cosh(r)-1-\sinh(r)\exp(i(\theta-2\phi))\}\right] = \hat{D}_1[\alpha\delta(e^{-r}-1)] \quad (3)$$

Notice that due to the mistake in sign [1] corrected in Eq. (1), the expression in (3) is different from that given in [1].

Substituting Eq. (2) and (3) into Eq. (1) and moving the squeeze operator to the left:

$$|\psi\rangle_{OUT} = \hat{D}_1\left(\alpha\delta(e^{-r}-1)\right)\hat{S}_1(\varsigma)\hat{D}_1(-\alpha\delta)\hat{D}_2(\alpha)|00\rangle =$$
$$\hat{S}_1(\varsigma)\hat{D}_1[\alpha\delta(e^{-r}-1)\cosh(r)+\alpha^*\delta(e^{-r}-1)\sinh(r)\exp(i\theta)]\hat{D}_1(-\alpha\delta)\hat{D}_2(\alpha)|00\rangle =$$
$$\hat{S}_1(\varsigma)\hat{D}_1[\alpha\delta(e^{-r}-1)\{\cosh(r)+\sinh r\exp(i(\theta-2\phi))\}\hat{D}_1(-\alpha\delta)\hat{D}_2(\alpha)|00\rangle$$
, (4)

and by using the condition $\theta - 2\phi = 0$, Eq. (4) is transformed to

$$|\psi\rangle_{OUT} = \hat{S}_1(\varsigma)\hat{D}_1[\alpha\delta(1-e^r)\hat{D}_1(-\alpha\delta)\hat{D}_2(\alpha)|00\rangle = \hat{S}_1(\varsigma)\hat{D}_1\left[-\alpha\delta e^r\right]\hat{D}_2(\alpha)|00\rangle \quad . \quad (5)$$

In both Refs. 1 and 2 the expectation values for the number operator $\langle\hat{n}\rangle$ and the standard deviation $(\Delta n)^2$ of a squeezed coherent state have been taken from Ref. (3), Eq. (2.7.13) and Eq. (3.5.18), respectively, which are given as:

$$\langle\hat{n}\rangle = |\alpha|^2\left(\cosh^2 r + \sinh^2 r\right) - 2|\alpha|^2\cos(\theta-2\phi)\sinh(r)\cosh(r) + \sin h^2 r \quad , (6)$$

$$(\Delta n)^2 = |\alpha|^2[\cosh(4r) - \cos(\theta-2\phi)\sinh(4r)] + 2\sinh^2 r\cosh^2 r \quad . \quad (7)$$

When we apply these equations to the present system we should insert the following changes: a) $|\alpha|^2$ should be exchanged into $|\alpha\delta|^2 e^{2r}$. b) The phase $\theta - 2\phi$ remains to be equal to 0 under the present condition. There is a subtle point which should be clarified here. $\alpha$ changes its sign on the right handside of Eq. (5) adding a phase $\pi$ to $\phi$ but this does not change the phase of $\theta - 2\phi$.

We get for the photon number expectation value :

$$\langle\hat{n}\rangle = |\alpha\delta|^2 e^{2r}\left[\left(\cosh^2 r + \sinh^2 r\right) - 2\cos(\theta-2\phi)\sinh(r)\cosh(r)\right] + \sin h^2 r \cong |\alpha\delta|^2 \quad , (8)$$

so that *the gravitational wave signal is not changed.* For $(\Delta n)^2$ we get

$$(\Delta n)^2 = |\alpha\delta|^2 e^{2r}[\cosh(4r) - \cos(\theta-2\phi)\sinh(4r)] + 2\sinh^2 r\cosh^2 r \cong |\alpha\delta|^2 e^{-2r} \quad (9)$$

so that under the condition $\theta - 2\phi = 0$ the quantum noise is reduced by a factor $e^{-2r}$. In Eqs. (8-9) we have neglected the small terms which are not proportional to the coherent light intensity. The problem in gravitational waves detection is that by increasing the light intensity for increasing the gravitational wave signal we increase the amount of noise. By using the present method for large values of $r$ ($r \geq 2$) the expectation value for $\langle \hat{n} \rangle$ can be increased by *increasing the light intensity* and at the same time getting smaller values of $(\Delta n)^2$ by having larger values of squeezing parameter $r$. One should notice that the quantum noise reduction is quite large for moderate values of $r$, e.g. for $r = 2$ approximately only 2% of the quantum noise remains. We should notice that if we substitute in Eq. (3) $\theta - 2\phi = \pi$ then the right side of Eq. (3) is changed to $\hat{D}_1[\alpha\delta(e^r - 1)$, which seems to represent amplification. However, due to the fact that certain phase relations are imposed in the present system between the $\hat{D}$ and $\hat{S}$ operators the full analysis does not show amplification *for a constant value of $\alpha$* [2]. One might eliminate the phases restriction by adding a BS in ouput port 1, and mixing the output squeezed state with another squeezed state but such analysis is beyond the scope of the present Note.

    I would like to clarify the following point: In both Refs. 1 and 2 the phase shift $\delta$ has been assumed to be real ,i.e., neglecting any phase difference between the gravitational phase perturbation and the coherent state. The gravitational wave is expected to increase one arm length of the interferometer and at the same time decrease the second arm length. During half time period of the gravitational wave this effect is reversed so that the tranverse time of the laser beam through the interferometer should be smaller than the time period of the gravitational wave. In a more general treatment than that made in [1] and [2] we should consider $\delta = |\delta|\exp(i\psi)$ where $|\delta|$ is the magnitude of the gravitational phase shift and $\psi$ will

represent its phase difference with the coherent state. The analysis under this general condition will remain almost the same but would change the phase difference difference $\theta - 2\phi$ to $\theta - 2\phi'$ where $\phi' = \phi + \psi$. Of course we cannot control the phase of the gravitational wave but we can change the phase of $\alpha$ so that the relation $\theta - 2\phi' = 0$ will be valid.

From a mathematical point of view we find that the unitary transformation of Eq. (5) in [1] on which both works [1,2] have been based can be changed into:

$$\begin{pmatrix} \hat{a}_1 \\ \hat{a}_2 \end{pmatrix} = \begin{pmatrix} \cos\gamma & \sin\gamma \exp(i\psi) \\ -\sin\gamma \exp(-i\psi) & \cos\gamma \end{pmatrix} \begin{pmatrix} \hat{b}_1 \\ \hat{b}_2 \end{pmatrix} \qquad , \qquad (10)$$

where $\psi$ is the phase difference between the gravitational wave and the coherent state. When the gravitational wave perturbation does not exist, then $\psi = 0$ and we return to the basic Eq. (5) of [1]. The gravitational wave leads to a change in the magnitude of $\sin\gamma$ like that which have been assumed in [1] and [2] but can also introduce the extra phase $\psi$. This does not change, however, the main features of the analyses as only phases differences are important.

The main conclusion from the present discussion is that the important effect which can be obtained in the present system is quantum noise reduction by the use of squeezed states and such conclusion is consistent with a previous analysis made by us [4].

**References**

1. R.Barak and Y.Ben-Aryeh, J. Opt. Soc. Am. B **25**, 361 (2008).

2. V.G. Voronov and M.Weyrauch, Phys. Rev. A **81**, 053816 (2010).

3. M.O. Scully and M.S. Zubairy, *Quantum Optics* (Cambridge Press, Cambridge, 1997 ).

4. O.Assaf and Y.Ben-Aryeh, J. Opt. Soc. Am. B **19**, 2716 (2002).